\documentclass[a4paper]{article}

\usepackage[english]{babel}
\usepackage[utf8]{inputenc}
\usepackage{amsmath}
\usepackage{graphicx}
\usepackage{fullpage}
\usepackage{setspace}
\usepackage{color}
\usepackage[font=footnotesize,compatibility=false]{caption}
\usepackage[font=footnotesize]{subcaption}
\usepackage{natbib}
\usepackage{authblk}
\usepackage{gensymb}
\usepackage{textcomp}

\usepackage{algorithm}
\usepackage{algpseudocode}

\floatstyle{plain}
\newfloat{myalgo}{tbhp}{mya}

{\begin{myalgo}[#1]
\centering
\begin{minipage}{0.8\textwidth}
\begin{algorithm}[H]}%
{\end{algorithm}
\end{minipage}
\end{myalgo}}

\begin{document}

\title{Single-Molecule Protein Identification by Sub-Nanopore Sensors}
\author[1]{Mikhail Kolmogorov}
\author[2]{Eamonn Kennedy}
\author[2]{Zhuxin Dong}
\author[2]{Gregory Timp}
\author[1]{Pavel Pevzner}
\affil[1]{Department of Computer Science and Engineering, University of California San Diego, USA}
\affil[2]{Electrical Engineering and Biological Science, University of Notre Dame, USA}

\maketitle

\abstract{
Recent advances in top-down mass spectrometry enabled 
identification of intact proteins, 
but this technology still faces challenges. For example, top-down mass 
spectrometry suffers from a lack of sensitivity since the ion counts for a single 
fragmentation event are often low.  In contrast, nanopore technology is exquisitely 
sensitive to single intact molecules, but it has only been successfully 
applied to DNA sequencing, so far. Here, we explore the potential of sub-nanopores
for single-molecule protein identification (SMPI) and describe an algorithm for 
identification of the electrical current blockade signal (nanospectrum) resulting
from the translocation of a denaturated, linearly charged protein through
a sub-nanopore. The analysis of identification p-values
suggests that the current technology is already sufficient for matching 
nanospectra against small protein databases, e.g., protein identification in bacterial proteomes.}

\section{Introduction}
When Church et al.~\cite{church1998measuring} proposed to use nanopores
for sequencing biopolymers, they had envisioned {\em both} DNA and proteins
sequencing. However, the progress in protein sequencing turned out
to be much slower since it is more difficult to force proteins through a
pore systematically and measure the resulting signal~\cite{timp2014think}.
These difficulties underlay the experimental and computational challenges of
{\em Single Molecule Protein Identification} ({\em SMPI}).

Nanopores promise single molecule sensitivity in the analysis of proteins,
but an approach for the identification of a single protein from its nanospectrum has remained elusive.
The most common approach to nanopore sequencing 
relies on the detection of the ionic–current blockade signal (nanospectrum) 
that develops when a molecule is driven through the pore by an electric field.
Preliminary work~\cite{meller2000rapid, sutherland2004structure}
was limited to analyzing protein conformations in pure solutions rather than identifying
proteins in a mixture. Subsequent steps demonstrated that nanopores
can detect protein phosphorylations~\cite{rosen2014single} as well as conformations  
and protein-ligand interactions~\cite{wu2014single}. Recent studies on combining
nanopores with aptamers have shown limited success for protein analysis~\cite{rotem2012protein}. 
Proposals  for electrolytic cell with tandem nanopores and for 
single molecule protein sequencing have been made, but not yet 
implemented~\cite{sampath2015digital,sampath2015peptide,sampath2015tandem,swaminathan2015theoretical}.

Recently, the sequence of amino acids in a denatured protein were read with 
limited resolution using a sub-nanometer-diameter pore, sputtered through a thin 
silicon nitride membrane~\cite{Kennedy2016}. When the denatured protein, immersed in electrolyte 
was driven through the pore by an electric field, measurements of a blockade 
in the current revealed regular fluctuations, the number of which coincides
with the number of residues in the protein. Furthermore, the amplitudes of the
fluctuations were correlated with the volumes that are occluded by quadromers 
(four residues) in the protein, but the correlation was imperfect, 
making it difficult to solve the problem of reconstructing a protein from its nanospectrum
with high fidelity. 

Developing computational and experimental methods 
for analyzing nanospectra derived from a electrical signals that produced when a 
protein translocates through a sub-nanopore could enable a real-time sensitive approach
to SMPI that may have advantages over top-down mass spectrometry for 
protein identification. Despite difficulty and expense
(requiring especially powerful magnets) to implement it, top-down mass spectrometry
has been used in a few labs around the world to identify intact proteins and their proteoforms. 
However, it is about 100-fold less sensitive than bottom-up mass spectrometry, 
which can be used to detect attomoles of material~\cite{pagnotti2011solvent}.
In stark contrast, a sub-nanopore has been used to discriminate residue substitutions 
in a \emph{single} 
%histone 
molecule with low fidelity~\cite{Dong2016}.

Similar to mass-spectrometry, where {\em de novo} protein
sequencing (based on top-down spectra) remains 
error-prone~\cite{liu2014novo,vyatkina2015novo}, the challenge of {\em de novo}
deconvoluting nanospectra into amino acids sequences of proteins is currently unsolved. 
However, protein identification based on top-down spectra 
(i.e., matching a spectrum against all proteins in a protein database) 
is a well-studied topic. For example, top-down protein identification tools 
ProsightPC~\cite{Zamdborg2007} and MS-Align+~\cite{Liu2013} reliably 
identify proteins, report p-values of resulting Protein-Spectrum Matches (PrSMs),
and even contribute to improving gene annotations by discovering previously
unknown proteins~\cite{kolmogorov2015spectrogene}.

In this paper, we describe the first algorithm for protein identification
based on nanospectra derived from current blockades associated with
denaturated, charge linearized translocation of protein through pores with
sub-nanometer diameters. Our Nano-Align algorithm matches nanospectra
against a protein database, identifies Protein-Nanospectrum Matches (PrNMs),
and reports their p-values. Our analysis revealed that the typical p-values
of identified PrNMs vary from $10^{-4}$ to $10^{-6}$, which
is already sufficient for a limited analysis of nanospectra against 
small bacterial proteomes.
%This computational analysis suggests further avenue for improving experimental protocols since
%improvement in the current fluctuations relative to the noise in a blockade
%promises to significantly reduce the p-values of PrNMs to make them commensurate with top-down proteomics. 

The software is publicly available at \emph{http://github.com/fenderglass/Nano-Align}.

% You may title this section "Methods" or "Models". 
% "Models" is not a valid title for PLoS ONE authors. However, PLoS ONE
% authors may use "Analysis" 
\section{Methods}

\subsection{Manufacturing sub-nanopores}  

\begin{figure}[ht]
	\centering
	\includegraphics[width=0.8\textwidth]{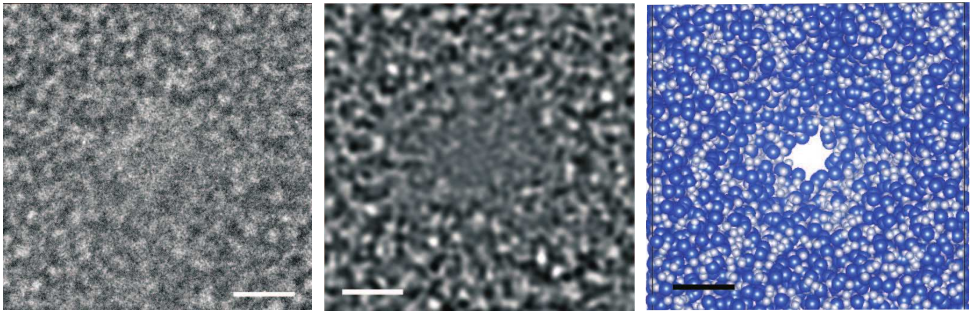}
	\caption{(left)
	TEM micrograph of sub-nanopore is shown with a 
	nominal diameter of 0.5 nm sputtered through silicon nitride membrane
	about 10-nm thick. The shot noise is associated with electron transmission
	through the pore. (center) Multi-slice simulations of the TEM image 
	are consistent with the experimental imaging conditions. The simulations
	correspond to a bi-conical pore with a 0.5 x 0.4 $nm^2$ cross-section and 
	a 15\degree cone angle at defocus of -40 nm. (right) Space-filled model 
	of the same pore is shown where the $Si$ atoms are represented by spheres 
	with a 0.235 nm diameter and $N$ atoms by spheres with a 0.13 nm diameter.
	The scale bars are 1 nm.}
\label{fig:pore_image}
\end{figure}

Pores with sub-nanometer cross-sections were sputtered through thin silicon
nitride membranes using a tightly focused, high-energy electron beam in a scanning
transmission electron microscope (Fig.~\ref{fig:pore_image})
as described in detail elsewhere~\cite{Kennedy2016}. 
The thickness of the membranes ranged from $8$ to $12$nm. 
%To discover the sub-nanopore topography, multi-slice algorithms were used to 
%calculate dynamic electron diffraction through model structures. 
%The model was partitioned into 40 equidistant slices along $z$-axis. 
%Phase gratings of the slices were calculated on grids with 512x512 pixels 
%in $x$ and $y$ for 300 keV incident electrons using the elastic and absorptive
%form factors and Debye-Waller factors to account for the thermal motion of the atoms.
%The calculations yielded an exit-plane wave function consistent with 
%the specified model structure.
%The simulations indicated that the pores were generally bi-conical,
%with cone-angles that ranged around $15 \degree \pm 5 \degree$, 
%and irregular, with sub-nanometer cross-sections at the waist.  
%The correspondence with the simulations ensured an accurate interpretation of the actual TEM image.

The silicon chip supporting a single membrane with a single sub-nanopore through it was bonded
to a polydimethylsiloxane microfluidic device
formed using a mold-casting technique. 
Two separate $Ag/AgCl$ electrodes 
were embedded in each of the microfluidic channels to independently electrically address
the cis and trans-sides of the membrane. To perform current measurements,
a sub-nanopore was immersed in 200-300 mM $NaCl$ electrolyte,  
a transmembrane voltage was applied
using $Ag/AgCl$ electrodes and the corresponding pore current was measured
using an Axopatch 200B amplifier.
Clampex 10.2 software was used for data 
acquisition and identifying regions of interest for further analysis.

\subsection{Signal acquisition} 

To measure a blockade current, a bias ranging from $-0.3V$ to $-1V$ was applied 
to the reservoir (containing 75 $\mu$L of electrolytic solution and 75 $\mu$L of 2x
concentrated solution of protein and denaturant, corresponding to about 20 fmoles of protein)
relative to ground in the channel. 
Lower bias improved the signal-to-noise ratio (SNR) of the data and lengthened the translocation times, 
but was observed to increase the probability of pore clogs and reduced the blockade rate. 
The background noise level was typically 12 pA-rms in 250 mM $NaCl$ solution at $-0.7V$.
Recombinant, carrier-free protein was reconstituted 
at high (100 $\mu$g/ml) concentration in PBS without adding BSA
to avoid false readings. From this solution, aliquots diluted to 2x the concentration
of denaturant with 200-500 pM protein, 20-100 $\mu$M BME, 400 mM $NaCl$ with $2 - 5 \cdot 10^{-3}$ 
\% SDS were vortexed and heated to 85\degree C for up to two hours. 
This concentration of NaCl electrolyte was chosen to screen the inherent pore charge
and avoid excess non-specific open pore noise. 
The protein solution was allowed to cool and added in 1:1 proportion 
with the (75 $\mu$L) electrolyte in the reservoir.
The low molarity of protein solution reduced the possibility that multiple molecules
compete for the pore at the same time. However multi-level events typically associated with residual 
native protein structure were still observed, but were manually culled from the data 
pre-analysis~\cite{Kennedy2016}. Data was recorded in 3 minute-long acquisition windows.

Five proteins
 were analyzed by measuring the blockade currents through sub-nanopores: 
a recombinant chemokine  CCL5 of length 68 AAs; two variants of the H3 histone designated as H3.2 and H3.3, 
which consist of the chain of 136 AAs, differing only by residue substitutions at positions 32, 88, 90 and 91;
a tail peptide of the H3 histone (residues 1-20) and a fourth histone, H4 of length 103 AAs.
More details about the datasets are given at the `Datasets' section below.

\subsection{Signal pre-processing}

\begin{figure}[ht]
	\centering
	\includegraphics[width=0.8\textwidth]{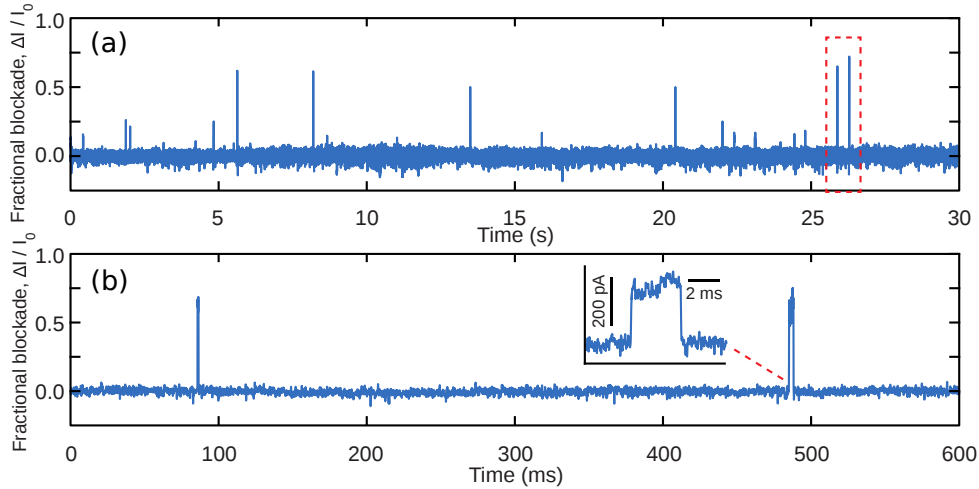}
	\caption{(a) An example of a pore current trace acquired from a denatured H3.3
	histone 
%in 250 mM NaCl and 0.01\% SDS 
translocating through sub-nanopore
	with a nominal diameter of 0.5-nm.
% with a 700 mV transmembrane bias applied.
	(b) The bottom trace is a magnified view of a 600 ms region of a top trace,
	showing a current blockade associated with the translocation of a single protein molecule.
	In the figure, higher values correspond to larger blockade currents. 
	Blockades, associated with the translocation of single proteins 
	were identified as regions with fluctuations five standard deviations above the noise level
	and with duration $>$ 1 ms. }
\label{fig:blockades_extraction}
\end{figure}

When a single molecule of protein translocates through the sub-nanopore, 
its amino acids block the flow of
ions, causing a change in the open pore current $I_{open}$. The fraction of occupied pore
volume $V_{mol} / V_{pore}$ (where $V_{pore}$ and $V_{mol}$
are volumes of the pore and molecule inside this pore, respectively) was assumed to be 
proportional to the {\em fractional blockade current}, 
which is calculated as $|I - I_{open}| / I_{open}$, where $I$ is the raw current during the translocation. 
The raw signal measurements from the pore were pre-processed as follows: first,
the discretized pore signal, sampled at 250 kHz, was split
into the separate blockades, each one representing a translocation of a single protein
(Fig.~\ref{fig:blockades_extraction}); and then the raw current $I$ was converted 
into \emph{fractional blockade current}.

Only events with sufficient duration to detect single-AA duration features were selected.
Typical blockade duration analyzed here ranged from 1 to 20 milliseconds,
as shorter times did not permit accurate discrimination of intra-event features
due to the measurement bandwidth. 
The mean fractional blockade current varied from 0.05 to 0.5 for different 
nanospectra. Recorded signals exhibited fluctuations that were 
associated with different structural features of a protein translocating through the pore.

\subsection{Mean Volume model of protein translocation}

Since the electrolytic current through the pore is associated with the
occupied pore volume, one of the major factors that influences the signal
is the volume of amino acids that occupy the sub-nanopore near 
the waist~\cite{coulter1953means}.
The estimates of amino acid volumes were obtained 
from crystallography data~\cite{perkins1986protein}.
Since the pore can simultaneously accommodate multiple amino acids, it was assumed that the
fluctuations in a blockade were proportional to a linear combination of amino acids volumes
in the pore waist. In particular, we found that the mean volume of amino acids 
yielded a good approximation of the empirical signal values.
Thus, given a protein $P$ of length $|P|$, we split it into overlapping windows
of size $k$ (or $k$-mers) and generate a {\em theoretical nanospectrum} $MV(P)$ as a vector  of
dimension  $|P| + k - 1$ by taking the average volume of $|P| - k + 1$ $k$-mers
and extra $2 * (k - 1)$ shorter prefix and suffix substrings from the beginning and end of a protein.
These extra prefix and suffix substrings correspond to the start and the end of a translocation, when the
pore is occupied by less than $k$ amino acids. 
For example, for $k=3$, the ``protein" \emph{KLMNP}  results in a
vector of length seven corresponding to the following substrings: \emph{K, KL, KLM, LMN, MNP, NP},
and \emph{P}. 

Experimental analysis of peptides with post-translational modifications~\cite{Kennedy2016}
and mutations~\cite{Dong2016} revealed changes
in the specific regions of the recorded signal traces, that corresponded to approximately
four amino acids in length. In addition, simulations of the electric field 
in a 0.5x0.5 $nm^2$ diameter, 8 $nm$ thick
pore in an \emph{SiN} membrane indicated that the vast majority of the field was confined 
within 1.5 nm of the pore near the waist at the center of the membrane, 
which gives roughly the same estimate of the number of amino acids.
Thus, the Mean Volume (MV) model assumes that each fluctuation in the blockade 
current corresponds to a read of a quadromer (short prefixes and suffixes of a protein 
correspond to shorter mers), which results in the best fit (among all reasonable values of $k$)
with experimental nanospectra.

%\subsection*{Reduced amino acid alphabet} 

%As many of the 20 proteinogenic amino acids have similar volumes, it can be difficult to distinguish
%between them based on the signal value in a nanospectrum because of the low signal-to-noise ratio. 
%We therefore partitioned the twenty amino acids into
%four groups according to their volumes (Fig.~\ref{fig:aa_volumes}):
%\textbf{L}arge ($> 0.2 nm^3$), \textbf{I}ntermediate (between 0.15 and 0.2 $nm^3$),
%\textbf{S}mall (between $0.11$ and $0.15 nm^3$) and \textbf{M}inuscule ($< 0.11 nm^3$).

\begin{figure}[ht!]
	\centering
	\includegraphics[width=\textwidth]{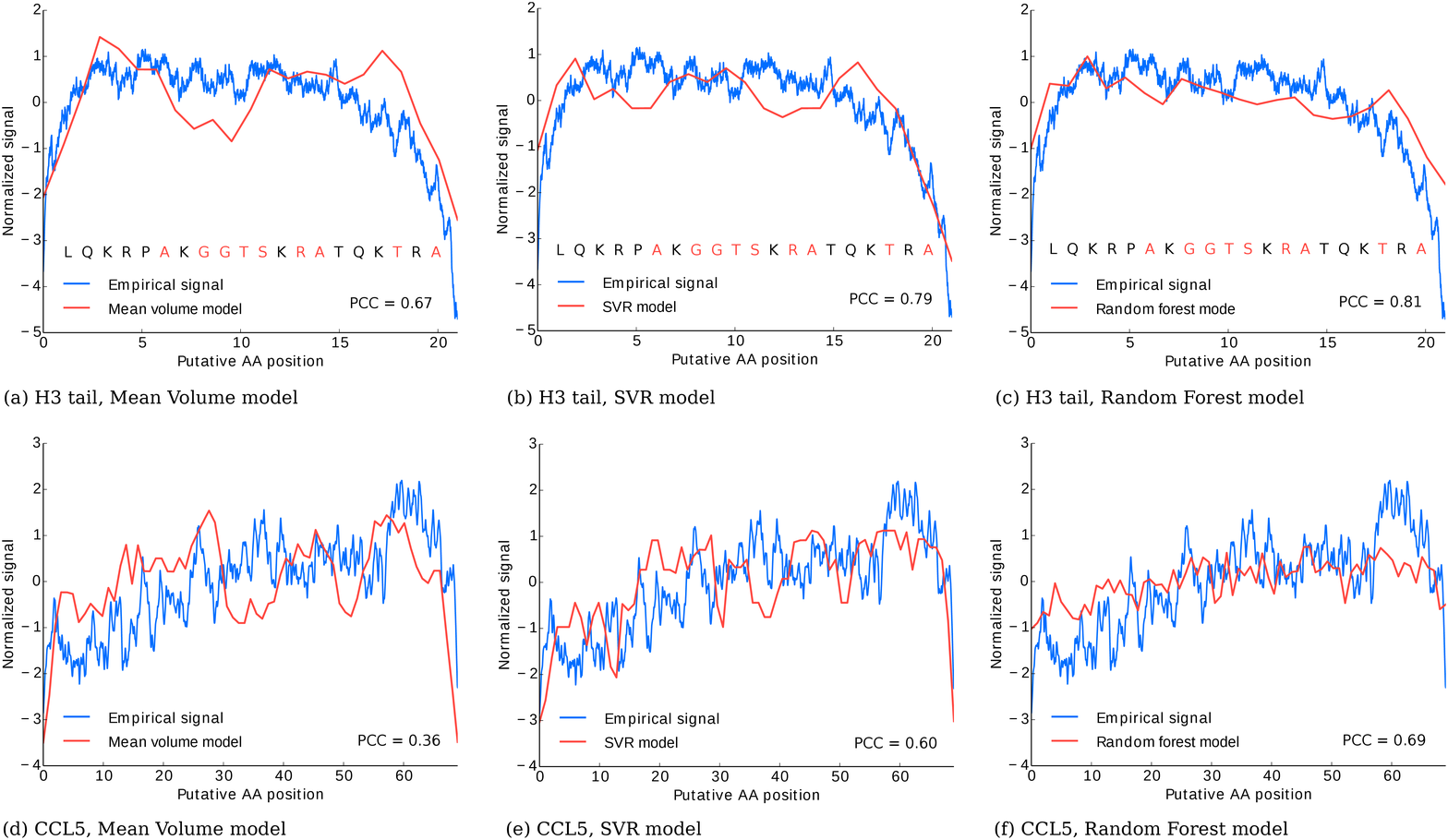}
	\vspace{3mm}

\caption{(a) A comparison of a consensus of 10 nanospectra of H3 tail peptide (10 AAs)
		and the corresponding theoretical nanospectra in the MV model ($k = 4$). 
		As the coefficients of the linear dependence 
		(for current vs. mean value) are unknown, each trace 
		was normalized by subtracting the mean and dividing them by the
		corresponding standard deviation. Each signal position is associated
		with an amino acid under the assumption that protein translocation
		velocity is uniform. The poorly-matched regions
		are enriched with smaller amino acids (with volumes below the median, 
		marked in red).
		The same comparison for the SVR model (b) and the RF model (c) shows better
		fit (measured as Pearson correlation coefficient). 
		Similarly, comparison of a consensus of 10 nanospectra of CCL5 peptide (68 AAs) 
		versus the MV, SVR, and RF models (d-f) shows
		improvement of the SVR and RF models over the MV model.}
\label{fig:models_example}
\end{figure}

%{\tt Should we generate a figure for hydrophilicity similar to Figure 4 and have them next to each other?} 

\subsection{Support Vector Regression-based model of protein translocation}

Generally, the MV model results in theoretical nanospectra 
correlated with the empirical data. The mean Pearson product-moment correlation coefficient
between a consensus of experimental nanospectra
(an average of multiple protein translocations, as described below)
and the corresponding MV model was ranging from $0.25$ to $0.45$ for various 
datasets. However some regions show large deviations
between theoretical and experimental nanospectra, which
may be associated with additional  attributes such as hydrophilicity  or charge. 
In particular, our analysis revealed that such discordant
regions were enriched with small amino acids, which have volumes below the median value
(see Fig.~\ref{fig:models_example} for illustration and `Characterizing errors of the models' 
section below for the detailed discussion).
Since we acquired multiple nanospectra originating from multiple known proteins,
an alternative approach for generating theoretical nanospectra was 
to use a supervised learning paradigm. We used a {\em Support Vector Regression} 
(an SVM-based regressor) to establish the correspondence between a $k$-mer inside
the pore and a signal it generates~\cite{scholkopf2001learning}.
Given an empirical nanospectrum $E$ recorded from a protein $P$,
we tiled $P$ into overlapping quadromers $q_i$ and discretized
$E$ into $|P| + 3$ points. Thus, each $q_i$ had an associated experimental signal value $e_i$.

\begin{figure}[ht!]
	\centering
	\includegraphics[width=0.3\textwidth]{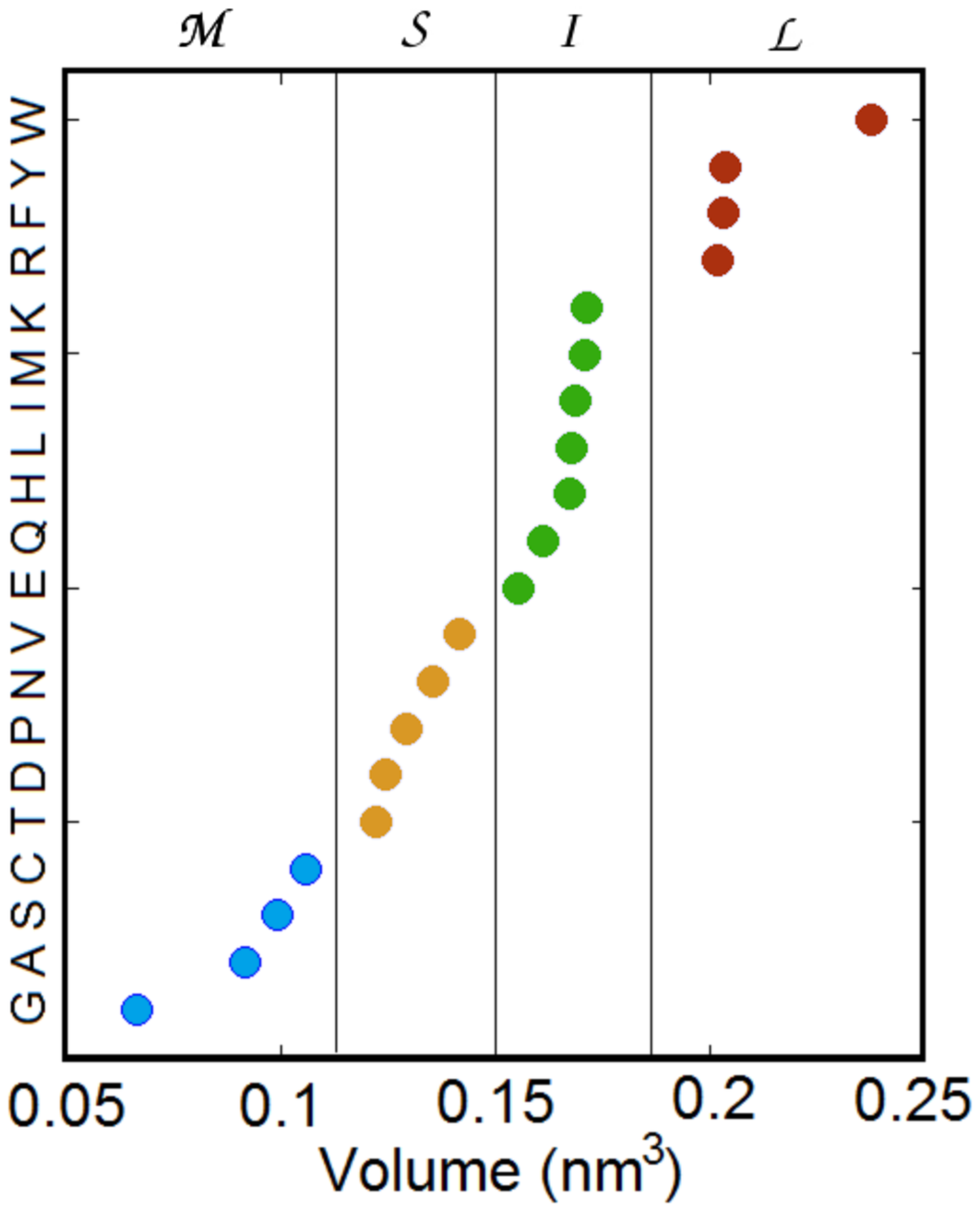}
	\vspace{3mm}

	\caption{Amino acids separated into four categories based on their volume: 
			 G, A, S, C ($\mathcal{M}$iniscule), T, D, P, N, V
			 ($\mathcal{S}$mall), E, Q, H, L, I, M, K ($\mathcal{I}$ntermediate), 
			 and R, F, Y, W ($\mathcal{L}$arge).}
	\label{fig:volume}
\end{figure}

Next, the feature space of the model has to be defined. Following the ideas of the MV model, 
it is natural to assume that blockade current is affected by the composition of
amino acids in a quadromer, rather than their order (however, the dependence might be non-linear).
As many of the 20 proteinogenic amino acids have similar volumes, 
%it can be difficult to distinguish
%between them based on the signal value in a nanospectrum because 
%(and the read fidelity is only about 0.07 $nm^3$). 
%Instead, 
we partitioned them into four volume groups
(Fig.~\ref{fig:volume}) and defined a {\em feature vector} $f_i$ of a quadromer $q_i$ 
as the composition of amino acids from each group (as a tuple of length four).
For example, an amino acid quardromer \emph{GQLD} has zero amino acids from 
$\mathcal{L}$arge group ($> 0.2 nm^3$), two from $\mathcal{I}$ntermediate group (between 0.15 and 0.2 $nm^3$),
one from $\mathcal{S}$mall group (between $0.11$ and $0.15 nm^3$) 
and one from $\mathcal{M}$inuscule group
($< 0.11 nm^3$), and is converted to a feature vector $(0, 2, 1, 1)$.
This choice of the feature space reduced
the overfitting effect and increased coverage of the training dataset
(there are only 35 distinct quadromer compositions in the defined feature space
versus $20^4$=160 000 amino acid quadromers).

Using a set of pairs $(f_i, e_i)$ we trained an SVR regressor with the Radial Basis Function
kernel (implemented in an open-source library libsvm~\cite{CC01a}).
The {\em Support Vector Regression (SVR)} model takes a peptide $P$ as input and outputs an SVR-based
theoretical nanospectrum $SVR(P)$ (Fig.~\ref{fig:models_example}).
The mean Pearson correlation coefficient between
the theoretical and empirical nanospectra (consensus)
for the SVRmodel was varying from $0.38$ to $0.68$ for different datasets,
confirming the improvement over the MV model.
The parameters of the SVR model were chosen through cross validation experiments
and are equal to $C=1000$, $\gamma = 0.001$, and $\epsilon = 0.01$.

\subsection{Random Forest-based model of protein translocation}

The analysis of error patterns of the SVR model revealed a bias in the signal
estimation that was correlated with the hydrophilicity of the amino acids
(see `Characterizing errors of the models' section).
Also, Bhattacharya et al.~\cite{bhattacharya2016water} recently reported that water molecules
affect the signal of DNA translocating through the nanopore since 
hydrophilic amino acids are more likely to acquire a water molecule and
change the effective volume~\cite{janin1979surface}. 
Thus, it is desirable  to include amino acid hydrophilicity into the model.

Motivated by these finding, we explored an alternative approach for supervised learning by using
the {\em Random Forest (RF) regression}~\cite{ho1995random,ho1998random} for
theoretical nanospectra generation. 
In comparison to the SVR model, the 
resulting {\em Random Forest (RF)} model is  more robust to outliers
and exhibit less overfitting~\cite{kleinberg1996overtraining}, which allowed us 
to use the volumes of all 20 amino acids as features.
According to this RF model, each quadromer $q_i$ 
from the training set is converted to a feature vector $f_i$,
where each element of the vector is a pair of volume and hydrophilicity
of the corresponding amino acid.

We used an open source implementation of the Random Forest regressor
from Scikit-learn package~\cite{scikit-learn} to build the described model.
The model performed well on the training sets, but the accuracy was
poor on the test proteins with different amino acid sequence and composition.
This was mainly caused by the fact that only a few among all possible amino acid
quadromers were observed in the training sets. However, under assumption that
nanopore current does not depend on the order of amino acids, it is possible to
significantly expand the training sets by randomly permuting amino 
acids within quadromers. Specifically, prior to model we randomly permuted each $f_i$ vector,
leaving the same corresponding $q_i$ value. This
dataset expansion significantly improved the performance of the RF model
on testing datasets. See Fig.~\ref{fig:models_example} 
for examples of theoretical nanospectra in the MV, SVR, and RF models.

\subsection{Protein identification}

Given an experimental nanospectrum $S$ and a protein $P$, we transformed $S$
into a vector $\vec{S}$ by splitting $S$ into $|P| + 3$ regions
and taking the average value inside each of them. The vector $\vec{S}$ was then normalized
by subtracting the mean and dividing by the standard deviation.
Under the hypothesis that $P$ has generated $\vec{S}$, we estimated the 
proportion of explained variance by computing $R^2$ coefficient of determination
between $\vec{S}$ and the model output.
Given a database of proteins $DB$, a protein $P(S,DB)$
is defined as a protein with the maximum $R^2$ against $S$
among all proteins from $DB$. A pair formed by the protein $P(S,DB)$ and the  
nanospectrum $S$ defines a putative {\em Protein-Nanospectrum Match (PrNM).}

\begin{figure}[ht!]
	\centering
	\includegraphics[width=\textwidth]{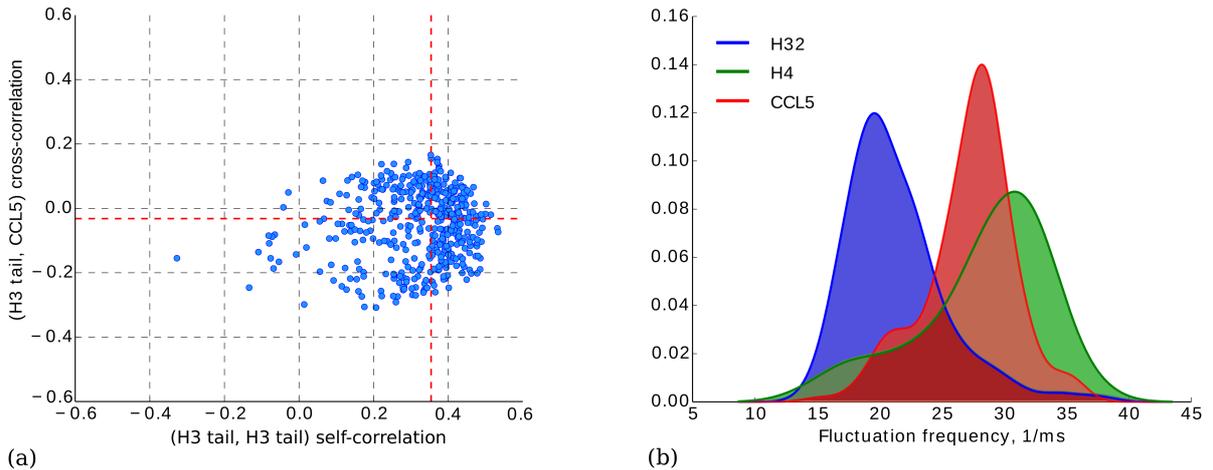}
	\vspace{3mm}

	\caption{(a) Cross-correlations compared with self-correlations between the
			 nanospectra originating from H3 tail protein and H4 protein. 
			 Cross-correlation values fluctuate around zero, while the median 
			 self-correlation is 0.35.
			 (b) Distributions of fluctuation frequencies reveals peaks
			 at positions 21 (H3.2), 34 (H4) and 29 (CCL5).}
	\label{fig:corr_noise}
\end{figure}

\subsection{Clustering nanospectra}

Single protein correlation analysis indicated that proteins were correlated 
more with themselves on average (Fig.~\ref{fig:corr_noise}a). 
In contrast, we did not observe such correlation in the open pore current, 
indicating that there is an inherent signal in blockades.
However, electrolytic current through
the pore is affected by many factors, such as uncorrelated time-dependent
fluctuations in the ionic current and electrical instrument noise,
which results in noisy nanospectra.
Averaging multiple nanospectra from the same protein resulted in
significant noise reduction and increased accuracy of PrNM identification. 
This effect is similar to improvements in peptide identifications that are achieved
by clustering of mass spectra in traditional
proteomics~\cite{frank2007clustering,frank2011spectral}.

Typically, clustering of $5-10$ nanospectra results in a consensus 
nanospectrum that significantly improves the signal-to-noise ratio 
over a single nanospectrum (the mean Pearson correlation coefficients between theoretical
and empirical nanospectra increased 1.5 -- 2-fold for various datasets). 
Since each of the existing datasets of nanospectra originated from a single pure protein,
we randomly partitioned the dataset of nanospectra
into clusters and performed identification of consensus nanospectra
instead of a single nanospectrum.

\subsection{Estimating protein length based on a blockade signal}

In traditional proteomics, the precursor mass assists top-down protein identification
since it greatly reduces the computational space
that has to be searched  in the protein database. Likewise, information about the protein length
would be very useful for SMPI, but estimating the protein length based on a nanospectrum
originating from a sub-nanopore is a non-trivial problem since the existing 
experimental protocol does not control
the translocation speed that may vary widely as evident from the blockade duration.

Our analysis revealed that protein translocations modulate the blockade current,
which was captured by the measurements.
Each blockade, associated with the translocation of a protein showed a characteristic
number of fluctuations during the duration of the blockade. It turned out that the fluctuation
frequency (described below) was correlated with the protein length and the other features,
such as amino acid composition.

We explored a possibility of the separation of a sample of nanospectra
into clusters corresponding to different proteins. 
From a sample of different proteins, we estimated the {\em fluctuation frequency} of
each nanospectrum  as the number of peaks (local maximums) divided by the duration of
the blockade. 
The distribution of fluctuation frequencies
(Fig.~\ref{fig:corr_noise}b) revealed that each protein in our datasets 
has a characteristic peak in the distribution.
To separate the nanospectra
into clusters based on the fluctuation frequency one can apply the Gaussian Mixture model to 
estimate the protein lengths from nanospectra and to improve the efficiency of  SMPI.

\subsection{The challenge of analyzing protein mixtures}

Analyzing a mixture of multiple proteins is conceptually
harder than analyzing the existing experimental datasets of nanospectra that all
originated from pure protein solutions. Since it is unknown what protein gives
rise to what nanospectrum in a mixture,
it is difficult to cluster nanospectra  for a reliable identification.
Further, orientation of each molecule must be deduced prior to clustering
since each protein can translocate through the pore in two different directions. 

However, it is possible to cluster nanospectra based on their estimated
fluctuation frequency  to differentiate proteins with different lengths.
As multiple proteins may have a similar length, it is important to further
split some \emph{length-based clusters} into finer \emph{protein-based clusters}.
We believe, that this could be done by applying clustering algorithms
which automatically estimate the number of clusters 
(e.g. Affinity Propagation~\cite{frey2007clustering}).
Evaluating the results of clustering in the case of
complex mixtures was problematic since all available experimental datasets of nanospectra
were generated from the pure protein solutions.

\section{Results}

\subsection{Datasets}

We benchmarked Nano-Align using nanospectra from  five short human proteins: H3.2, H3.3, H4, CCL5
and H3 tail peptide (Table~\ref{tabular:datasets}).
The nanospectra from H3.2, H3.3 and H4 were acquired using the two similar pores
whereas the nanospectra for CCL5 and H3 tail were acquired using two different 
pores with different sizes. 
The proteins were split into three pairs: (CCL5, H3 tail), (H4, H3.2) and
(H3.3, H3.2). For each pair of proteins, the SVR and RF models were trained using the 
protein with higher number of nanospectra and the accuracy of identifications was estimated 
using the other protein from the pair.
The first two pairs represented proteins that were very different in both length and amino acid composition,
thus minimizing the overfitting effect. The third pair represented
highly similar proteins, that only differ in four amino acids.

\begin{table}[h]
  	\begin{center}
 	\begin{tabular}{c c c c c}
	\hline
	 Dataset & Peptide length & \#Nanospectra & Pore id & Pore size (nm$^2$)\\
    \hline
	H3.2 & 136 & 445 & ZD349, ZD350 & 0.6x0.5\\
	H3.3 & 136 & 25 &  ZD349 & 0.6x0.5\\
	H4 & 103 & 89 & ZD350 & 0.6x0.5 \\
	CCL5 & 68 & 239 & ZD158 & 0.8x0.6 \\
	H3 tail & 20 & 477 & ZD220 & 0.6x0.5 \\
   \hline
  \end{tabular}
  \caption{Datasets summary.}
  \label{tabular:datasets}
  \end{center}
\end{table}

\subsection{Evaluating protein identification accuracy}

To evaluate the accuracy of SMPI, we constructed decoy protein database
for each dataset from the correct protein and randomly generated
proteins of the same length and amino acid composition as the correct protein.
The size of  decoy database varied from $10^5$ to $5 \cdot 10^6$ for different
datasets, 
%{\tt Why not to unify it and make it $10^7$ for all spectra?} 
depending on the identification accuracy and the number of nanospectra
in the dataset.
The p-value of a PrNM was approximated as the percentage of proteins
from the database scoring higher than the correct protein 
against the given nanospectrum.

\begin{figure}[ht!]
	\centering
	\includegraphics[width=\textwidth]{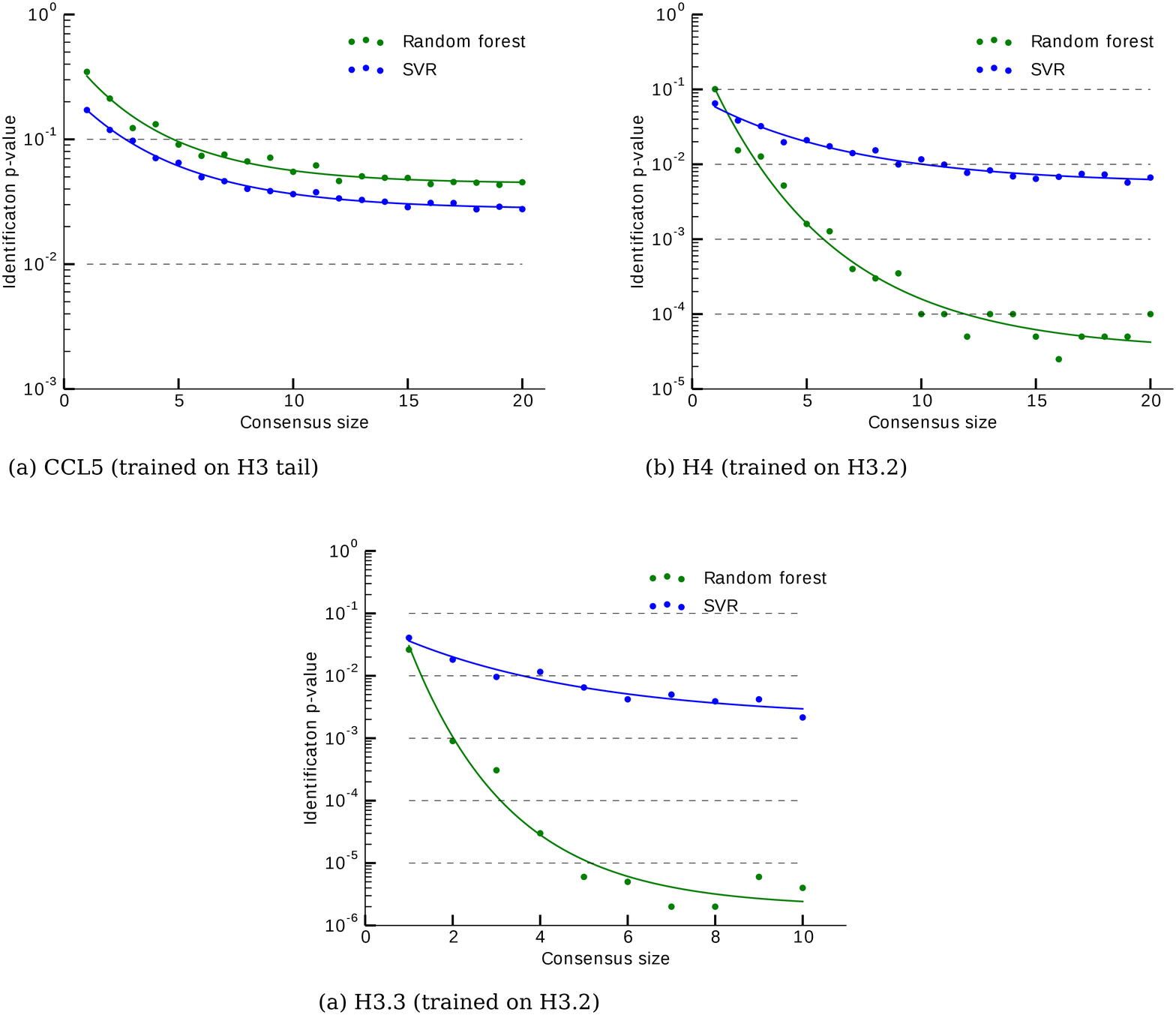}
	\vspace{3mm}

\caption{Median p-value as a function of the number of nanospectra in a cluster (fitted with
		 the exponential curve). Decoy database sizes are $10^5$ for H3 tail and H4 datasets
		 and $5 \cdot 10^6$ for H3.3 dataset. Significant
		 outliers with respect to the fitting curve were filtered out. The median p-value
		 of the RF model on the H3.3 dataset turns into zero for the consensus size exceeding 10.}
\label{fig:error_rates}
\end{figure}

Below we show results for the SVR and RF models only since they 
turned out to be significantly more accurate than the MV model for all datasets.
Fig.~\ref{fig:error_rates} shows median p-values for 
SVR and RF models as a function of the number of nanospectra in a cluster.
As expected, both models showed the improvement in the accuracy with
the increase in the cluster size. The p-values for the pair (CCL5, H3 tail) were
high for both models (0.03 - 0.05 for a consensus of size 10). However,
the dataset (H4, H3.2) showed a significant improvement for the RF model
(p-values of the order of $10^{-4}$ for a consensus of 10 nanospectra),
while the accuracy of the SVR model was comparable to the previous dataset.
Finally, the RF model showed high accuracy on (H3.3, H3.2) dataset, 
with p-values below $10^{-5}$ for the nanospectra clusters of size five.

The RF model consistently outperformed the SVR model on the datasets that were generated
using pores of similar sizes, which suggests that the decision trees are better suited
for SMPI due to their robustness against outliers. Also,
amino acid hydrophilicity proved to be a valuable predictor of the pore signal.
The RF model performed slightly worse than the SVR model on the dataset generated 
using two different pores, suggesting that 
it is more sensitive to the experimental conditions.
The fact that the RF model performed better on the proteins that were more similar
to the training proteins is not surprising, but rather highlights the importance 
of choice of the training set, which should have substantial coverage of the
data.

Additionally, we benchmarked the RF model performance using a database containing 
real human proteins. We extracted all 
proteins of length between 100 and 160 from the human proteome (about 20\% of the
human proteome) and performed the identification of H3.3 spectra against this
reduced database. On average, the true protein was ranked five against all
other proteins (for a cluster of size five). An example of database
hits is given in the Table~\ref{tabular:h33_human}. Interestingly, all high-scoring
proteins belong to H3 histone family and differ by only few amino acids.
While the search space was artificially reduced, this experiment already provides
a justification for analysis of unknown nanospectra against small bacterial proteomes,
after further improvements in the protein length estimation discussed above.

\begin{table}[ht]
  	\begin{center}
 	\begin{tabular}{c c c c}
	\hline
	 Rank & Protein Id & $R^2$ score & Length \\
    \hline
	1 & H3F3B & 0.4002 & 132 \\
	2 & HIST2H3A & 0.3989 & 136 \\
	3 & HIST1H3A & 0.3980 & 136 \\
	4 & H3F3C & 0.3905 & 135 \\
	\textbf{5} & \textbf{H3F3A} & \textbf{0.3871} & \textbf{136} \\
	6 & HIST3H3 & 0.3819 & 136 \\
	7 & HIST2H3PS2 & 0.3714 & 136 \\
	8 & PRR14 & 0.3248 & 104 \\
	9 & BRD8 & 0.3146 & 122 \\
	10 & ANAPC16 & 0.3028 & 110 \\
   \hline
	\vspace{3mm}
  \end{tabular}
  \caption{An example of H3.3 nanospectra identification (for a cluster of size five) against
  		   all human proteins of length 100 - 160 AAs. The total database size is 
		   14 293, which covers approximately 20\% of the human proteome taken from the UniProt database. 
		   The correct protein is shown
		   in bold. Proteins from the H3 family exhibit the highest $R^2$ scores among other proteins
		   from the database.}
  \label{tabular:h33_human}
  \end{center}
\end{table}

\subsection{Characterizing errors of the models}

\begin{figure}[ht!]
	\centering
	\includegraphics[width=\textwidth]{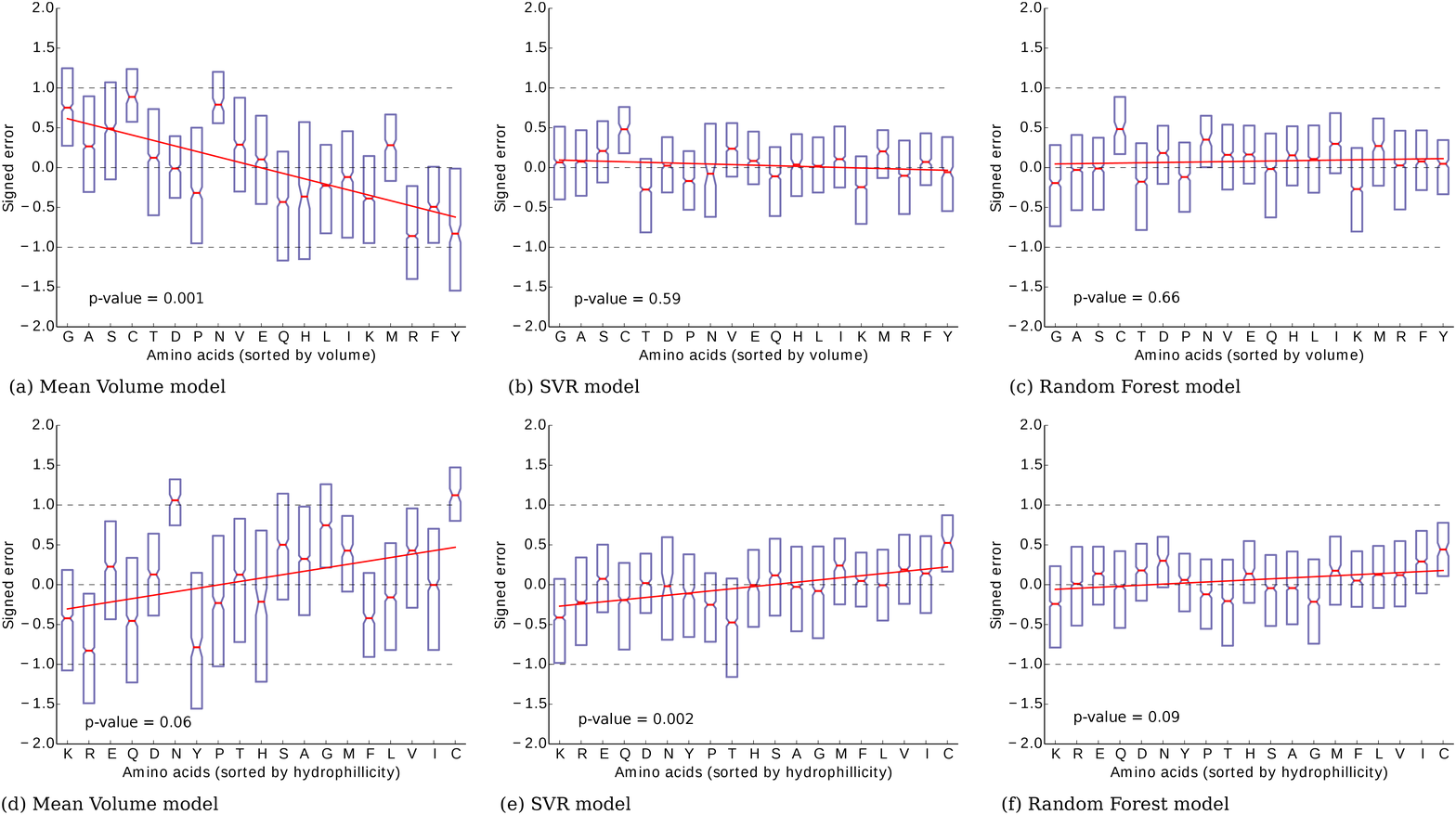}
	\vspace{1mm}

\caption{ Error (calculated as the difference between empirical and
		theoretical nanospectra) for amino acids of H3.2 protein sorted in the increasing 
		order of their volumes. (a) MV model has a tendency to underestimate signal
		associated with small amino acids and overestimate signal from large amino acids.
		SVR model (b) and RF model (c) trained on H3.2 dataset do not have volume-related bias.
		As \emph{Trp} is not present in H3.2, it is not shown on the figure.
		P-values are given for the hypothesis that linear slope is non-zero.
		Similar analysis reveal signal bias with respect to amino acid hydrophilicity
		for SVR model (e). MV model (d) and RF model (f)
		do not show statistically significant bias.}
\label{fig:model_bias}
\end{figure}

For each of the three models (MV, SVR and RF) we measured the bias with respect to different
features of amino acids. Using H3.2 dataset (that provides the best amino acid coverage) 
we calculated  the \emph{signed error} defined as the
mean difference between the empirical and theoretical nanospectra. For each amino acid,
the signed error was measured among the associated quadromers.
Fig.~\ref{fig:model_bias} shows the volume-related bias of the MV model.
This bias could be explained by the fact that larger amino acids have more
influence on the pore signal than smaller amino acids.
The SVR model and RF model show no bias with respect to amino acid volumes.
A similar analysis revealed a bias with respect to amino acid hydrophilicity
in the SVR model. The MV model did not show a clear dependence, possibly due to the
dominant effect of the volume bias. The RF model showed no statistically
significant bias related to hydrophilicity.

\section{Discussion}
We presented the first algorithm for Single Molecule Protein
Identification  using a signal generated by a protein translocation
through a sub-nanopore. We also proposed three models for generating theoretical
nanospectra and concluded that the Random Forest model results in the
most accurate identifications. The typical estimated p-values of identification accuracy
were ranging from $10^{-4}$ to $10^{-6}$, which is already sufficient for a 
limited analysis of nanospectra against small bacterial proteomes containing a few thousands proteins.
The comparison of algorithm performance on different datasets suggests that 
the model sensitivity will further improve when more nanospectra originated
from different proteins become available.

Cysteine (Cys) was the highest source of error in all three models for H3.2.
Likewise, Cys was an above average source of error in CCL5~\cite{Kennedy2016} but, 
it was a below average source of error in the similar sequence of CXCL1. 
Thus, it seemed unlikely that only the size affects the  error. 
On the other hand, both Cys and Met, which exhibit higher number of prediction 
errors are at the high end of the hydropathy index and have only few waters (4 and 10, respectively) 
binding them~\cite{thanki1988distributions}, which may indicate that water affects
the blockade current. In addition, it has been speculated that charge could also 
affect the duration and magnitude of a blockade~\cite{Kennedy2016,kowalczyk2012slowing}.
Whereas it seems likely that both charge and water play a role in the blockade current, 
measurements and the MV model testing these ideas have been inconclusive so far~\cite{Dong2016}. 

While SMPI is currently not in a position to compete with top-down proteomics, 
this technology is still in its infancy. 
Furthermore, due to the inherent single molecule sensitivity, 
there are several avenues of research that can be addressed
uniquely by SMPI that offer protein-discrimination from very small samples (attomoles).
Thus, SMPI has a potential to emerge as a new technology  
for accurate protein identification.

\bibliography{mybib}{}
\bibliographystyle{plainnat}

\end{document}